\documentclass[12pt,a4paper]{article}
\pdfoutput=1
\usepackage[utf8]{inputenc}
\usepackage{amsmath}
\usepackage{amsfonts}
\usepackage{amssymb}
\usepackage{graphicx}
\usepackage{cite}
\author{Xianglin Liu \\
{\small \textit{Department of physics, Carnegie Mellon University} }\\
Yang Wang \\
{\small \textit{Pittsburgh supercomputing center, Carnegie Mellon University}}\\
Markus Eisenbach  \\
{\small \textit{Center for computational sciences, Oak Ridge National Laboratory} }\\
G. Malcolm Stocks \\
{\small \textit{Materials science and technology division, Oak Ridge National Laboratory} }
}

\newcommand{\mr}{\mathbf{r}}

\newcommand{\ud}{\underline}
\begin{document}
\title{Relativistic full-potential multiple scattering theory: An \textit{ab initio} method and its applications}
\maketitle

\begin{abstract}
The Green function plays an essential role in the Kohn-Korringa-Rostocker (KKR) multiple scattering method. In practice, it is constructed from the regular and irregular solutions of the local Kohn-Sham equation and robust methods exist for spherical potentials. However, when applied to a non-spherical potential, numerical errors from the irregular solutions give rise to pathological behaviors of the charge density at small radius. Here we present a full-potential implementation of the relativistic KKR method to perform \textit{ab initio} self-consistent calculation by directly solving the Dirac differential equations. The pathology around the origin is completely eliminated with the help of an efficient pole-searching technique. This method is utilized to investigate the crystal structures of polonium and their bulk properties. The noble metals are also calculated, both as a test of our method and to study the relativistic effects.
\end{abstract}

\section{Introduction} \label{introduction}
Multiple-scattering theory (MST) underpins a number of widely used methods for solving the electronic structure problem in periodic solids, all of which have their origins in the KKR method originally introduced by Korringa\cite{Korringa} in 1947 and independently re-derived by Kohn and Rostoker\cite{Kohn} in 1953.  Two features of MST distinguish it from conventional Raleigh-Ritz variational approaches. Firstly,  it naturally yields a separation between the single-site potential scattering and structural arrangement (positions) of the individual scatterers.
Secondly,  in the framework of Density functional theory (DFT), it provides an explicit expression for the Green function of the system, which can then be used to calculate the charge and spin densities without explicit calculation of the wavefunctions that are the focus of Raleigh-Ritz methods. The availability of the Green function makes MST a versatile tool that can be easily combined with other methods to investigate more complex systems. For example, by applying the Dyson's series expansion to the Green function, defects and impurities in an otherwise perfect crystal can be investigated \cite{Defects}. Another example is the KKR-CPA method, which is based on a combination of MST with the coherent potential approximation (CPA) \cite{Soven, Gyorffy, StocksCPA} to calculate the configurationally averaged properties of disordered systems, such as random alloys \cite{JohnsonPRL, JohnsonPRB} and the disordered local moment state of metallic magnets \cite{DLM}.
A more recent development is in studying of strongly correlated systems, where the MST Green function can be readily used in conjunction with the GW approximation\cite{GW} or the dynamical mean field theory (DMFT) \cite{DMFT}. Moreover, the real space formulation of MST \cite{LSMS} has demonstrated essentially ideal linear scalability on current supercomputing architecture\cite{Markus}, and, as a result, can be employed to study solid state systems with tens of thousands of atoms.

The originally formulated MST solved the Schr\"odinger equation within the muffin-tin (MT) potential approximation, where the potential is assumed to be spherically symmetric within the muffin-tin spheres and constant in the interstitial region \cite{ Korringa, Kohn}. While the muffin-tin approximation generally works well for systems dominated by metallic bonding, it cannot properly describe a wide range of systems where the asymmetries of the effective potential \cite{KohnSham} play an important role,
such as surfaces, two-dimensional materials, and systems with directional covalent bonding.  In addition, because the Schr\"odinger equation is nonrelativistic, it cannot properly describe systems where relativistic effects are important. In particular, it doesn't account for the spin-orbit coupling (SOC), a subject currently of great interest due to its role in many technically important phenomen, such as magnetocrystalline  anisotropy, Rashba effect, and magnetic Skyrmions\cite{Skyrmion}. To take into account the relativistic effects, a common practice is to treat the relativistic kinematic effects with the scalar-relativistic approximation \cite{Koelling}, and include the SOC in a  perturbative second-variational way. However, this strategy could be problematic for heavy-element systems where SOC is not small compared crystal field splitting. To take into account both relativity and the full shape dependence of the crystal potential on an equal footing, the original MST formulation must be extended to a full-potential, Dirac equation, based theory, and indeed much work has been done in this regard by a number of groups\cite{Tamura, Lovatt, XDWang, Huhne, Geilhufe, RLS, Xianglin}.

In MST the Green function is constructed from the regular and irregular solutions of the Kohn-Sham equations. In contrast to the MT scheme, a persistent problem in standard implementation of full-potential MST is that numerical errors in the irregular solutions are very difficult to control near the origin\cite{Rusanu}. As a result, the charge density calculated from the Green function exhibits pathologies which can extend to a sizable fraction of the muffin-tin radius. The practice employed by Huhne \textit{et al} \cite{Huhne} is to drop the non-spherical components of the  potential within a cutoff radius $r_{ns}$, with the argument that close to the nucleus the non-spherical contributions to the potential are small.
However, for an accurate determination of the Hellmann-Feynman forces, these non-spherical components of the potential are important and cannot simply be discarded. In Ref. \cite{Akai}, a modified single-site Green function is proposed
to avoid directly using the irregular solutions, which, however, still show up a volume integration in the expression of the Green function.
In Ref. \cite{Zeller} it is proposed to use a sub-interval technique to systematically decrease the numerical error by reducing the step size when approaching the origin. This method still requires spherical potential approximation within a small radius, and is also less effective for larger $l_{max}$, which signifies the angular momentum cutoff of the solutions.

To completely get rid of this pathology, we use an approach developed from the work of Rusanu \textit{et. al.}\cite{Rusanu}, where the use of irregular solution is avoided by splitting the full Green function into two parts and carrying out the energy integration along different contours on the complex plane. One major difference between our method and the one in \cite{Rusanu} is in the energy integration of the single-site Green function, where the presence of the single-site resonance states and shallow bound states make a naive integration scheme laborious on the real energy axis. By making use of an efficient pole-searching algorithm of the authors, Y. Wang, we find this real axis energy integral can be accomplished much faster. Furthermore, because it explicitly identifies both bound and virtual bound electron states, our method provides an excellent framework for implementing schemes, such as LDA+U \cite{Vladimir} and self-interaction correction (SIC) \cite{Perdew, Luders}, aimed at correcting local approximations to the DFT for the effects of strong correlation.

In the following section we will first explain the scheme that is used for performing contour integration of the MST Green function, then we will show how the poles of the single-site Green function can be used to facilitate the energy integration of the shallow bound states and the resonance states. Details of this pole-searching technique are presented in the appendix. In section \ref{polonium}, polonium is used as an example to demonstrate our method. The lattice constants, bulk modulus and crystal structures of Po are studied and compared with results from other methods. In section \ref{noble metals}, the density of states and bulk properties of copper, silver, and gold are calculated as a further test of our method and to quantify the increasing impact of relativistic effects.

\section{Methods} \label{methods}
The two physical quantities of most interest in the present context are the integrated density of states $N(E)$ and the charge density $\rho(\mr)$. In a typical \textit{ab initio} DFT calculation, these quantities need to be evaluated at each self-consistent loop to determine the new Fermi energy and effective potential. In MST, the charge density is obtained from the energy integral of the Green function,
\begin{align}
\rho(\mr)=-\frac{1}{\pi} \mathrm{Im} \; \mathrm{Tr}\int_{E_b}^{E_F} G(E,\mr,\mr) dE, \label{ChargeDensity}
\end{align} 
where $E_b$ is the bottom of the conduction band, $E_F$ is the Fermi Energy. The integrated density of states (IDOS) is given by the energy integral of the density of states (DOS) $n(E)$
\begin{align}
N(E)=\int_{E_b}^{E} n(E') dE',
\end{align}
 and $n(E)$ is calculated from the volume integral of the Green function
\begin{align}
n(E)=-\frac{1}{\pi} \mathrm{Im} \; \mathrm{Tr}\int_{\Omega} G(E,\mr,\mr) d\mr. \label{DOS}
\end{align}
The Green function is obtained by solving the Dirac-Kohn-Sham equation and the details have been given in Ref. \cite{Xianglin}. Note that the preceding equations involve an energy integration along the real energy axis in order to obtain $N(E)$ and $\rho(\mr)$. Unfortunately, for bulk materials a simple energy integration on the real axis turns out to be infeasible due to the dense set of poles in the corresponding multiple scattering Green function. One resolution of this problem is to carry out the integral along a contour in the complex energy plane \cite{ContourInt}, with the observation that the Green function is holomorphic except for poles at the bound states and a cut on the real axis starting at $E_b$. Because the DOS becomes increasingly smooth the further the contour is distorted into the complex plane, this method has been found to be very efficient. Indeed, deploying Gaussian quadrature integration method, only a few dozen energy points are needed to reach a high accuracy. In practice, however, implementation is hindered by the presence of the irregular solutions in the expression of the Green function, which is commonly written as \cite{Xianglin}
\begin{align}
G(E, \mr,\mr')=&\sum_{\Lambda \Lambda'}Z_\Lambda(E, \mr) \tau_{\Lambda\Lambda'}(E) Z^+_{\Lambda'}(E, \mr')-\sum_{\Lambda} Z_\Lambda(E, \mr) \mathcal{J}^+_{\Lambda}(E, \mr'). \label{Green function}
\end{align}
where $\Lambda$ stands for the pair of relativistic angular-momentum indices ($\kappa$,$\mu$), $Z_\Lambda(E, \mr)$ and $Z^+_\Lambda(E, \mr)$ are the right-hand and left-hand regular solutions, respectively,  $\mathcal{J}^+_{\Lambda}(E, \mr')$ are the left-hand irregular solutions, and $\tau_{\Lambda\Lambda'}(E)$ are the scattering-path matrices. The irregular solutions are singular at the origin and are obtained by integrating inward from outside of the bounding sphere. Unfortunately, using standard numerical integration algorithms, the irregular solutions typically have unacceptable numerical errors near the origin, which then results in the aforementioned pathology in the charge density. . The reason for this instability is as follows: In full-potential scheme, the non-spherical potential components couple solutions of the differential equation having different $l$ indices. Near the origin, the irregular solutions diverge as $r^{-l-1}$, and the coupling of this divergence to that of channels of higher $l$ amplifies the numerical round-off error in the irregular solutions; an effect that is further amplified as the $l_{max}$ use in the differential equation solver is increased. 

In our approach, the elimination of this pathological behavior of the Green function is accomplished by spitting the Green function into the single-site scattering part $G_{s}$ and the remaining multiple scattering part $G_{m}$, as suggested in Ref. \cite{Rusanu},
\begin{align}
G(E, \mr,\mr')
=&G_{s}(E, \mr,\mr')+G_{m}(E, \mr,\mr') .
\end{align}
The explicit expression of the $G_{s}(E, \mr,\mr')$ is
\begin{align}
G_{s}(E, \mr,\mr')=&\sum_{\Lambda \Lambda'}Z_\Lambda(E, \mr) t_{\Lambda\Lambda'}(E) Z^+_{\Lambda'}(E, \mr')-\sum_{\Lambda} Z_\Lambda(E, \mr) \mathcal{J}^+_{\Lambda}(E, \mr'), \label{Singel-site Green}
\end{align}
where $t_{\Lambda\Lambda'}(E)$ is the single-site $t$ matrix. For simplicity of the discussion, we only consider the case of one atom per unit cell. The explicit expression of $G_{m}(E, \mr,\mr')$ is
\begin{align}
G_{m}(E, \mr,\mr') =&\sum_{\Lambda \Lambda'}Z_\Lambda(E, \mr) \left( \tau_{\Lambda\Lambda'}(E)  - t_{\Lambda\Lambda'}\right) Z^+_{\Lambda'}(E, \mr'). 
\label{Green} 
\end{align}
Note that the $Z\mathcal{J}$ term in $G_{s}(E, \mr,\mr')$ is real for real energies. Because the single-site DOS and charge density involve only the imaginary part of $G_{s}(E, \mr,\mr')$ this term can be ignored. As a result, carrying out the required integration over real energy axis obviates the need to evaluate irregular solutions. As for the multiple scattering contribution, because both $G_{s}(E, \mr,\mr')$ and $G(E, \mr,\mr')$ are holomorphic on the upper half-plane, it can be easily integrated along a semi-circle contour, as shown in Fig. \ref{fig:Contour}.

To efficiently evaluate the energy integral of $G_{s}(E, \mr,\mr')$, two obstacles must be overcome. The first one is to properly account for the contribution from sharp resonance states, examples of which are the $d$-state resonances in noble metals \cite{Xianglin} and the $p$-state resonances of polonium, as shown in Fig. \ref{fig:DOS_Po}. In relativistic full potential schemes, these resonance peaks usually get sharper and split into multiple peaks due to spin-orbit coupling and crystal field splitting, and make a straightforward energy integration even more prohibitive. The second difficulty is to carry out the energy integration of the shallow bound states, which have poles right on real axis and make a direct numerical integration unfeasible. These shallow bound states show up, for example, in our calculation of polonium, as listed in Table {\ref{table:poles}.

The resonance peaks originate from the poles located at the forth quadrant in the complex plane. As will be explained in the following, these peaks are well approximated by the Lorentzians, and the energy integration on the positive axis can be carried out efficiently with a weighted sampling technique. To accomplish this, the single-site Green function in the neighborhood of a scattering resonance is first approximated as
\begin{align}
G_s(E,\mr,\mr)_{|E\approx E_n} \approx \frac{\psi_n(E,\mr)\psi_n^{\dagger}(E,\mr)}{E-(E_n-i\lambda_n)}, \label{spectral green}
\end{align}
where the complex resonance energy has be written in terms of the real and imaginary part as $E_n-i\;\lambda_n$. By substitution of equation (\ref{spectral green}) into equation (\ref{DOS}) and using the normalization condition of the wave functions 
\begin{align}
\int_{\Omega} \psi_n(\mr) \psi^{\dagger}_n(\mr) d\mr=1,
\end{align}
then the density of states around $E_n$ becomes
\begin{align}
n(E)_{|E\approx E_n}
\approx &-\frac{1}{\pi} {\rm{Im}} \left(\frac{1}{E-E_n+i\lambda_n}\right) = \frac{1}{\pi} \frac{\lambda_n}{(E-E_n)^2+\lambda_n^2}, \label{peaks}
\end{align}
which is exactly a Lorentzian function. The values of $E_n$ and $\lambda_n$ are determined using the pole-searching technique detailed in section \ref{pole}. Now that the approximate behavior of the resonance peaks are known, we can construct a weighted energy mesh to carry out the integration, i.e., an energy mesh that is densest at the resonance peaks. To use this method, we need to find an appropriate cumulative distribution function $F(E)$. Here it is chosen to be 
\begin{align}
F(E)=&\sum_n \left(\frac{1}{\pi}\rm{arctan}\left( \frac{E-E_n}{\lambda_n}\right)+\frac{1}{2}\right)+\frac{V}{3\pi^2} E^{3/2} ,
\end{align}
where the first part is the integral of the Lorentzian function and the second part is to account for the non-resonance (free-electron) states, with V being the volume of the unit cell. The weighted energy mesh is obtained by uniformly choosing points $F_i$ between $F(0)$ and $F(E_F)$, then solving the inverse of $F(E)$ with, for example, bisection method.

For shallow bound states, the above weight sampling technique no longer works because what was a Lorentzian function at positive energy evolves into a Dirac delta function at negative energies. We instead carry out the energy integration on small contours that encircle the negative energy poles ( see Fig. \ref{fig:Contour}). The reason that radius of the contour must be small is to reduce the error caused by not carrying out the integration strictly on real axis. In our experience, this method yields accurate results when the radius of contour chosen to be $10^{-4} $ Ry.

The numerical parameters used in the calculations are as follows: The angular momentum cut-off of the wave function is chosen to be $l_{max}=4$ and 4096 k-points are used in the first Brillouin zone unless otherwise specified. The number of energy points used in the weighted sampling integration is 100. Thirty Gaussian energy points distributed on a large semi-circular contour that encompasses the full valence band are used to integrate $G_m(E,\mr,\mr')$ and 5{\textendash}10 energy Gaussian points are used for the small contours around shallow bound states.

\begin{figure}[h!]
\centering
   \includegraphics[width=0.8\textwidth]{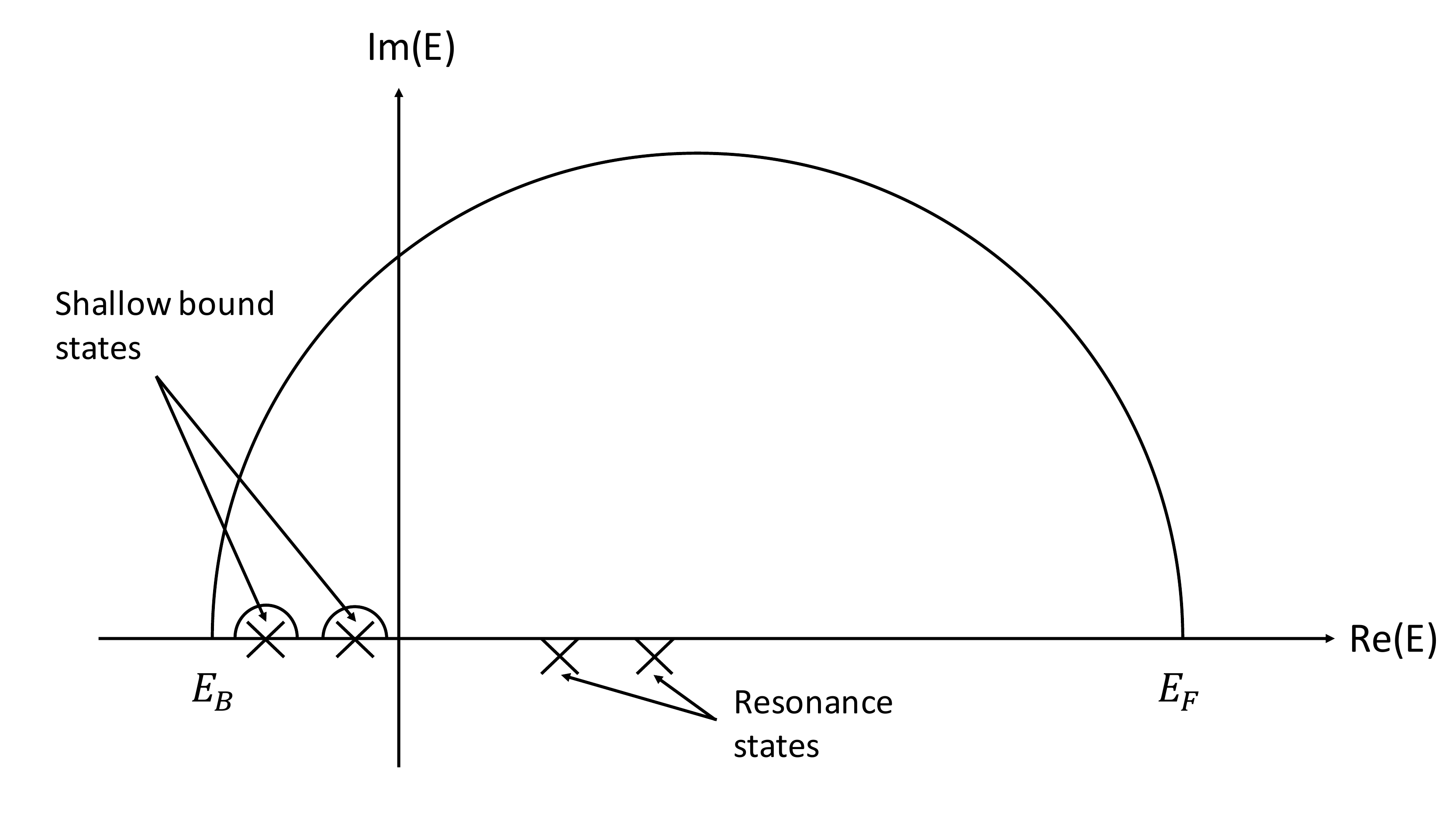}
\caption{  The Green function is split into two parts. The  multiple scattering part $G_{m}$ is integrated along the upper semi-circle contour, while the single-site part $G_{s}$ is integrated on real axis. The shallow bound states are integrated with a tiny circle and the resonance states are integrated using weighted sampling technique.   } \label{fig:Contour}
\end{figure}

\section{ Polonium } \label{polonium}
Polonium is an element that is extremely toxic and highly radioactive, and experimental data on its physical properties is scarce. One distinctive property of Po is that it is the only element to crystallize in the simple cubic (sc) structure at ambient conditions. The simple cubic structure has a low atomic packing factor, and is generally considered unstable, both from the point of views of elastic stability \cite{Elastic} and Peierls instability \cite{Min}.  In addition to the simple cubic phase Po ($\alpha$-Po), a second, slightly distorted, rhombohedral phase, Po ($\beta-$Po), is also found to exist at elevated temperatures. Both the two crystal structures of Po are studied by Beamer and Maxwell\cite{Beamer} using X-ray diffraction, and the lattice constants are reported to be $a=3.345(2)\,\mathrm{\AA}$ for $\alpha$-Po and $a=3.359(1)\, \mathrm{\AA}$ for $\beta$-Po. Later Sando and Lange\cite{Desando} redetermined the lattice constants using purer Po sample, with $a=3.359(1)\,\mathrm{\AA}$ for $\alpha$-Po and $a=3.368(1)\,\mathrm{\AA}$ for $\beta$-Po.

With the development of electronic structure calculation methods, a number of theoretical studies have been carried out to explain why Po has a stable sc structure. There are still debates on this question, but the general consensus is that relativistic effects play an important role. Using the pseudopotential (PP) method, Kraig \textit{et al.} \cite{Kraig} showed that sc structure has the lowest total energy among a number of lattice configurations, including face-centered cubic structure (fcc) and body-centered cubic structure (bcc).
Min \textit{et al.} \cite{Min,KimPRL} ultilized the full-potential linearized augmented plane wave (FLAPW) method implemented in the WIEN2K package and found the sc structure is due to SOC. Also using FLAPW, Legut \textit{et al.} \cite{LegutPRB, LegutPRL} investigated the total energy profile of Po and analyzed contributions from different relativistic terms, and concluded instead that it is the relativistic mass-velocity and Darwin terms that stabilize the sc structure. This controversy is mainly due to the tiny total energy difference (especially for all-electron calculation) between the sc structure and trigonal structure of Po, which is at the order of 0.1 mRy, and can be affected by factors such as accuracy of exchange-correlation functional, convergence of the total energy with respect to numerical parameters (for example, we found the convergence with respect to radial grid is important because for different structures the grid generated can be different.), or even which experimental value of the lattice constant to use in the band and phonon dispersion calculations \cite{MinPRB}.
Compared to the second-variational implementation of SOC in WIEN2K, the fact that our method includes the SOC intrinsically provides some advantage, but in this paper we do not intend to address the question on which relativistic  effects contribute more to the stabilization of sc-Po, mainly because of the aforementioned tiny total energy difference. Instead, we use Po as an example to demonstrate our method, and compare the calculated physical quantities with experiment and other calculation results. 

The electron configuration of Po is [Xe]$4f^{14}5d^{10}6s^26p^4$. The core and semi-core electrons are calculated by solving the Dirac equation for the spherical (l=0) component of  the potential. The valence electrons, i.e., $6s$ and $6p$ states, are calculated with the FP-MST method of this paper. As described in section \ref{methods}, the Green function is split into a single-site part and a multiple scattering part. The poles of the single-site Green function are computed and listed in Table \ref{table:poles}, and the indices in the first column signify the corresponding predominate angular momentum quantum numbers of each pole. The first two $s$ poles and the following two $p$ poles have negative real part and negligible imaginary part, which are the characteristics of bound states. The other four $p$ poles are in the forth quadrant of the complex plane and correspond to resonance states. The splitting of the six $p$ electrons into a doubly degenerate and weakly bound $p_{1/2}$ core state and a $p_{3/2}$ fourfold degenerate resonance state just above the energy zero is due to SOC. Note that the cubic symmetry in the potential does not break the degeneracy of either of these spin-orbit split manifolds. The spherical components of the charge densities of the $s_{1/2}$ and $p_{1/2}$  bound states are shown in Fig. \ref{fig:ChargeD_Po}. Clearly, the charge densities demonstrate correct number of nodes. Note that these nodes result from the orthogonality of the valence states to the core states and occur for a radius as small as 0.1 a.u.. As in Ref. \cite{Rusanu}, this is a further example where the previous method of simple polynomial extrapolation the charge density, over some significant fraction of the muffin-tin radius, will miss the effects of these undulations. 

The single-site DOS and the total DOS of the system are shown in Fig. \ref{fig:DOS_Po}; the latter is in good agreement with the scalar relativistic + SO interaction calculation result of Ref. \cite{LegutPRB}. The total energies and the fitted equation of state Po with sc, bcc and fcc structures are shown in Fig. \ref{fig:Po_fit}. The lattice constant for the sc phase is found to be $a=3.327$ {\AA}, which, as can be seen in Table \ref{table:Po}, is in good agreement with the experimental value $a=3.359$ {\AA}. The bulk modulus is determined to be $B_0=55.1$ GPa, but unfortunately no reliable experimental value of $B_0$ is known. In Ref. \cite{Lach}, it is claimed that the experimental bulk modulus of $\alpha$-Po is 26 GPa, but no reference is given. In Ref. \cite{Rubio} a more complete list of the theoretical results of $\alpha$-Po is given, from which we see all bulk moduli calculated theoretically are in the range of $40-60$ GPa, which is much larger than the 26 GPa quotation. We therefore conclude that the bulk modulus found here is reasonable and consistent with the values in the literature.

\begin{table}[ht]
\caption{The poles of the single-site Green function of Po }
\centering 
\begin{tabular}{l l l } 
\hline\hline 
  & real part & imaginary part  \\ [0.5ex] 
\hline 
$s_{1/2}$ & -0.68202472D+00 & 0.16325374D-15 \\ 
                & -0.68202472D+00 & -0.21372993D-15 \\
\hline
$p_{1/2}$ & -0.35205135D-01 & -0.12391130D-13 \\
                & -0.35205135D-01 &  0.11887440D-13  \\
\hline
$p_{3/2}$ & 0.66052385D-01 & -0.37929455D-01  \\
                & 0.66052385D-01 & -0.37929455D-01  \\ 
                & 0.66052385D-01 & -0.37929455D-01  \\ 
                & 0.66052385D-01 & -0.37929455D-01  \\  
\hline 
\end{tabular}
\label{table:poles} 
\end{table}

\begin{table}[ht]
\caption{Comparison of the lattice constant ($a$) and bulk modulus ($B_0$) calculated with the experimental and theoretical data in the literature. }
\centering 
\begin{tabular}{l l l l } 
\hline\hline 
  & a(\AA)  & $B_0$ (GPa) & Ref.  \\ [0.5ex] 
\hline 
Exp. & 3.345(2) &  &\cite{Beamer} \\ 
                & 3.359(1) &  &\cite{Desando}\\
LDA, PP  & 3.28 & 56  & \cite{Kraig}\\
GGA+SO, FLAPW & 3.34 &   & \cite{Min}\\
LDA+SO, FLAPW & 3.334 & 42.3 &\cite{LegutPRB} \\
LDA+SO, FLAPW & 3.323 & 47.35  & \cite{Rubio}\\
 TB+SO & 3.29& 51  & \cite{Lach}\\ 
LDA+Dirac, MST & 3.327 & 55.1  & This work\\  
\hline 
\end{tabular}
\label{table:Po} 
\end{table}

\begin{figure}[h!]
\centering
   \includegraphics[width=0.7\textwidth]{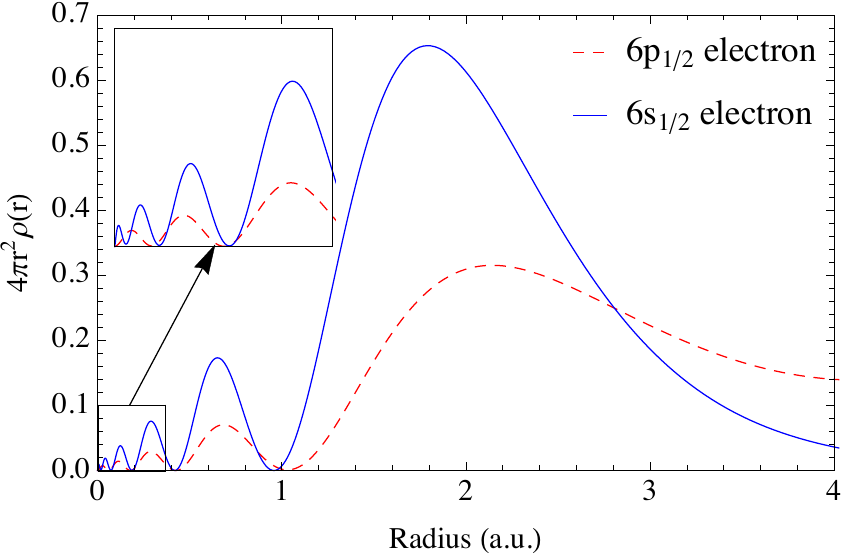}
\caption{ (Color online) The spherical component of the charge density of polonium corresponding to the shallow bound states. Note that a $4\pi r^2$ factor has been included. The \textit{red dashed} line corresponds to $6p_{1/2}$ electron and the \textit{blue solid} line corresponds to $6s_{1/2}$ electron. } \label{fig:ChargeD_Po}
\end{figure}

\begin{figure}[h!]
\centering
   \includegraphics[width=0.7\textwidth]{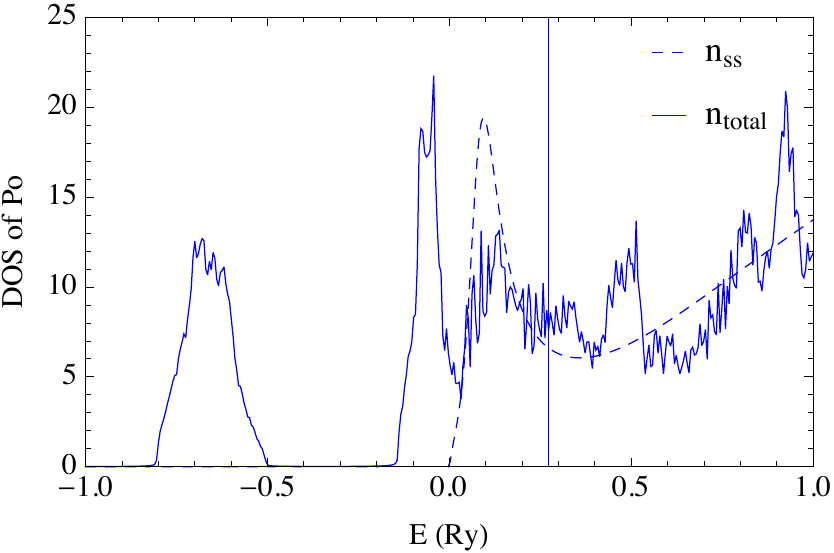}
\caption{ Comparison of the total DOS and the single-site DOS of polonium. To show the DOS, the energy is shifted a little up on the real axis, with imaginary part $\mathrm{Im(E)}$=0.001 Ry. The \textit{dashed} line is the single-site DOS and the \textit{solid} line is the total DOS. The vertical line is the Fermi energy. The shallow bound states are not shown on the single-site DOS because they are essentially a set of Dirac $\delta$ functions. 125000 k-points are used to calculate the DOS.} \label{fig:DOS_Po}
\end{figure}

\begin{figure}[h!]
\centering
   \includegraphics[width=0.7\textwidth]{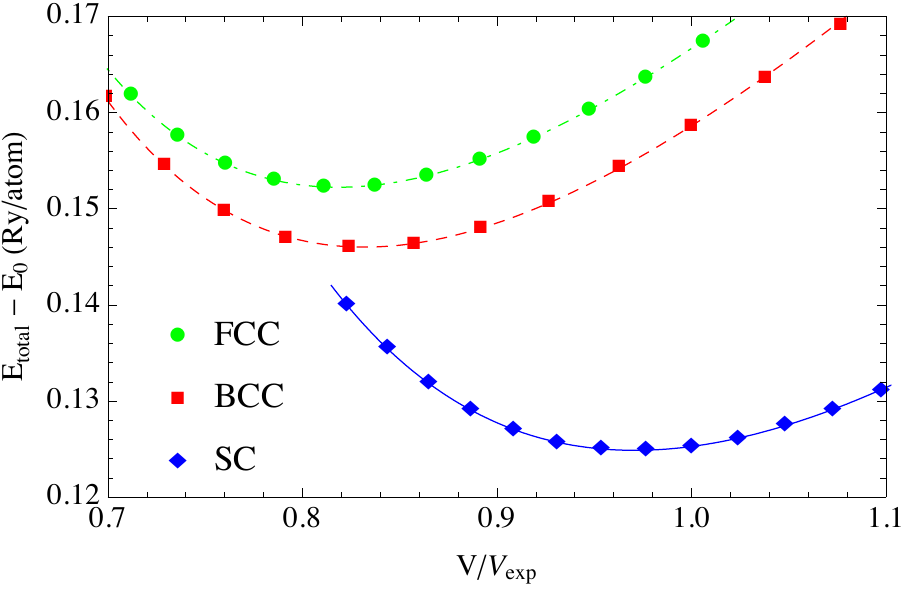}
\caption{ (Color online) The volume dependence of total energy plot of polonium. The \textit{solid}, \textit{dashed}, \textit{dot-dashed} lines correspond to the fitted equation of state of simple cubic, BCC, and FCC structures and the shift of total energy $\mathrm{E_0}=-41342$ (Ry/atom). When solving the Dirac equations a total of 1,500 radial mesh points are used within the muffin-tin radius to ensure convergence of the single-site solutions.} \label{fig:Po_fit}
\end{figure}

\section{ Noble metals} \label{noble metals}
\begin{table}[ht]
\caption{Comparison of calculated lattice constants and bulk moduli of noble metals with the experimental values. The LDA is employed in all calculations. Fully relativistic schemes are used except for the last two columns. In column $\rm{Au_{NR}}$ all electrons are calculated with nonrelativistic schemes, and relativistic schemes are utilized only  for valence electrons in column $\rm{Au_{R-{Core}}}$ .    }
\centering 
\begin{tabular}{c c c c c c } 
\hline\hline 
  & Cu & Ag & Au &$\rm{Au_{NR}}$ & $\rm{Au_{R-{Core}}}$ \\ [0.5ex] 
\hline 
 Lattice constants (a.u.) \\
 This work & 6.65 & 7.55 & 7.60 &8.03& 8.17 \\ 
Experiment & 6.84 & 7.72 & 7.71 & &
\vspace{4 mm}\\
 \hline
Bulk Modulus (GPa) \\
This work & 191 & 141 & 208 &112 & 124\\
Experiment & 137 & 101 & 180 & & \\ [1ex]  
\hline 
\end{tabular}
\label{tab:metals} 
\end{table}
In contrast to polonium, the noble metals have been thoroughly investigated both theoretically and experimentally. In this section, we use them as examples to further test our method. 
The lattice constants and bulk moduli calculated here are shown in table \ref{tab:metals} together with experimentally measured values. The results are in excellent agreement with the ones calculated with conventional KKR method \cite{Papanikolaou}. Compared to the experimental data, we underestimate the lattice constants and overestimate the bulk modulus, which is a well-known characteristic of LDA. As demonstrated in Ref. {\cite{Papanikolaou}, by employing GGA instead of the LDA functional, much better agreement with the experimental data can be obtained for noble metals. 

To demonstrate the impact of relativity, we calculated the bulk properties of Au with two other schemes. In the nonrelativistic (NR) scheme all relativistic effects are ignored, and in the relativistic core (R-Core) scheme relativistic effects are included only for the core and semi-core electrons. The lattice constants and bulk modulus are shown in the last two columns of Table \ref{tab:metals} and the total energy vs lattice constants plot is shown in Fig. \ref{fig:Au_fit}. From these results, it is clear that relativistic effects need to be included for all the electrons, core and valence, when calculating the ground state properties of elements as heavy as Au.

The impact of relativity is also evident in the DOS plots of Fig. \ref{fig:DOS_metals}. As a point of reference, the fully relativistic DOS are in good agreement with the results in Ref. \cite{MacDonald} for the same set of systems. As can be seen, in all nonrelativistic calculations, the DOS have five main peaks, which is a result of the crystal field symmetry. Under the increasing influence of spin-orbit coupling, these peaks further split into multiple sub-peaks. Overall, the differences between relativistic and nonrelativistic results also grow as the atomic number increase. For silver and gold, the $d$-bands obtained in the relativistic calculation broaden significantly with the result that the top of the $d$-band is much closer to the Fermi energy than in Cu. This effect is the result of the relativistic contraction of the inner shell $s$ electrons, whose relativistic effects are more significant than the $d$-electrons. Actually, this is the well-known explanation of the color of gold; namely, that relativistic effects, decrease the transition energy between $5d$ and $6s$ states which then absorb  blue light making the reflected light appear golden to us. The relativistic effects in silver are small compared to gold, therefore Ag still reflects all the visible wavelengths and appears “silver”. Concerning the energy difference between top of the $d-$band and the Fermi energy in the relativistic plots, we find they are small compared to those obtained from photoemission measurements\cite{Panaccione, Strocov}. Again, this is a typical feature of DFT calculations, and is shown to be largely corrected \cite{Marini} by using GW method to account for self-energy effects.

\begin{figure}[h!]
\centering
  \includegraphics[width=0.7\textwidth]{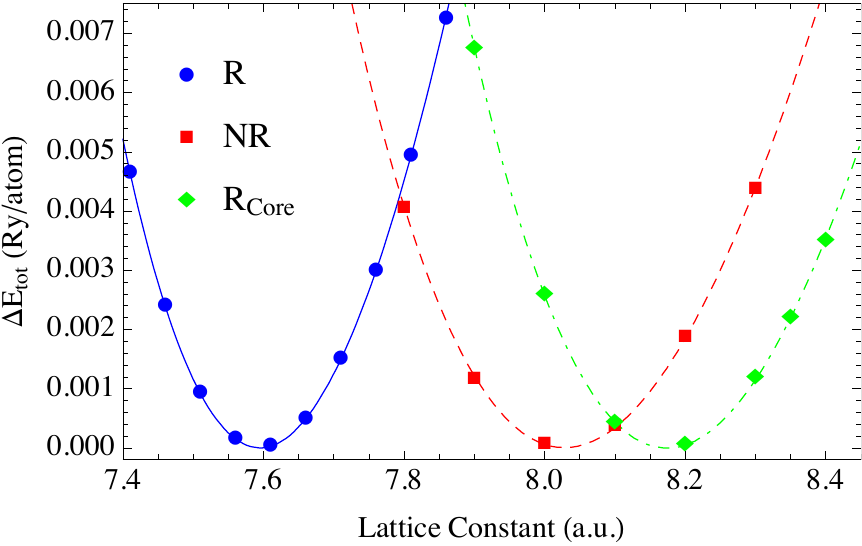}
\caption{(Color online) The total energy of Au calculated using different methods. Different energy shifts are made for each curve so that the lowest point of each line is always zero. In the \textit{solid} line, the core, semicore, and valence electrons are calculated by solving the Dirac equations. In the \textit{dashed} line, the nonrelativistic Schr{\"o}dinger equations are used for all the electrons. In the \textit{dot-dashed} line the Dirac equations are solved for valence electrons while the Schr{\"o}dinger equations are solved for core and semi-core electrons. } \label{fig:Au_fit}
\end{figure}

\begin{figure}[h!]
\centering
  \includegraphics[width=0.7\textwidth]{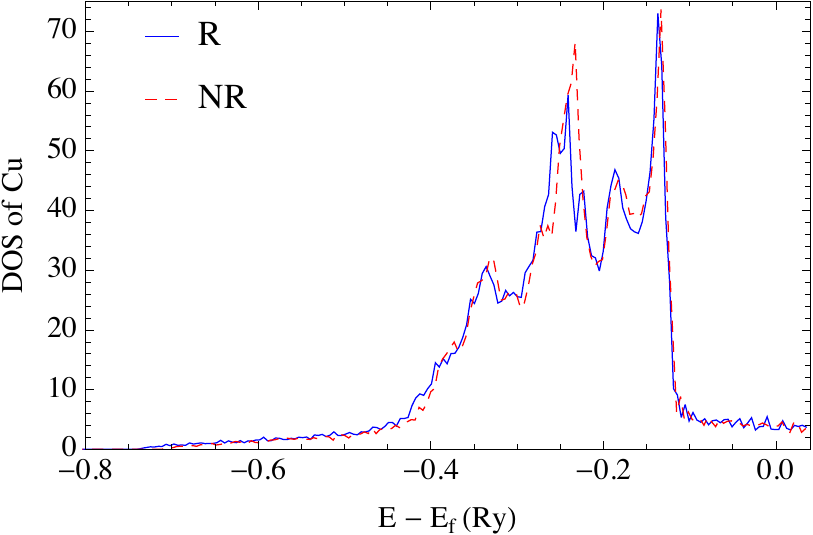}
  \includegraphics[width=0.7\textwidth]{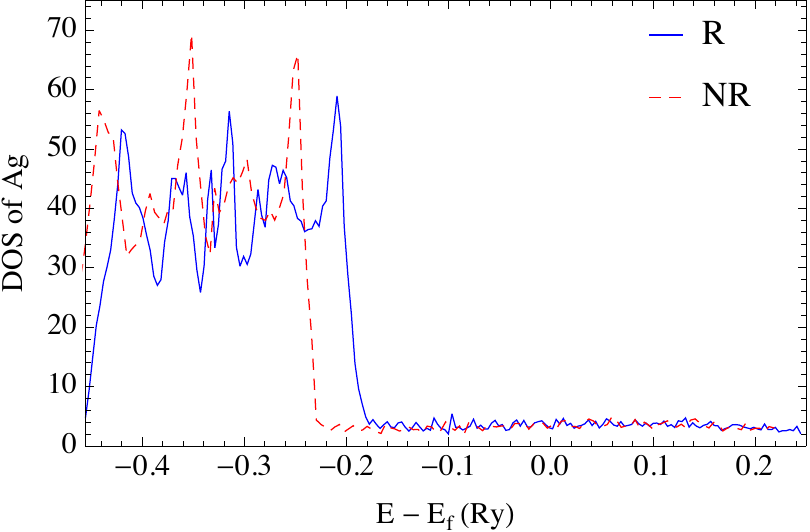}
  \includegraphics[width=0.7\textwidth]{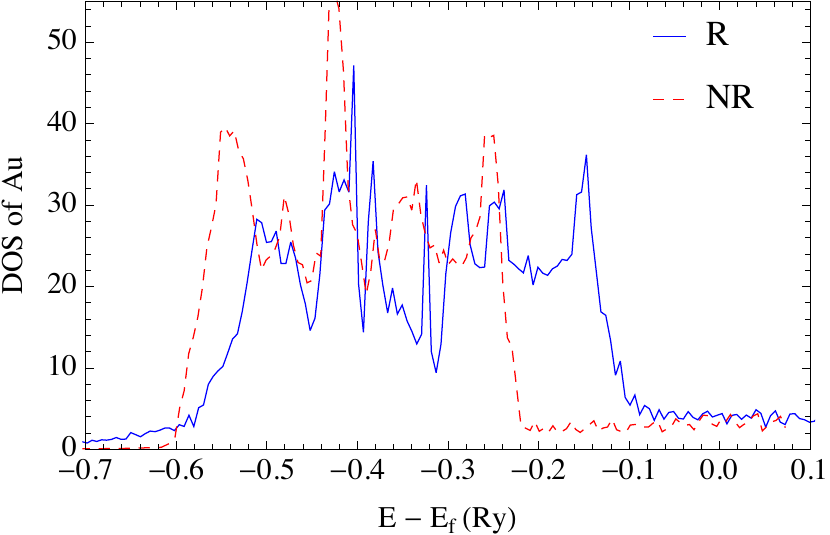}
\caption{(Color online) Comparison of the relativistic and nonrelativistic DOS of Cu, Ag, and Au. The \textit{red dashed} lines are the nonrelativistic results and \textit{blue solid} lines are the relativistic results. Note that shift of the energy by $\mathrm{E}_{\mathrm{f}}$ has been applied, so zero on the x-axis corresponds to the Fermi energy.}  \label{fig:DOS_metals}
\end{figure}
\section{Conclusions}
We have implemented a new approach to full-potential relativistic MST. By splitting the Green function into two parts and carrying out the energy integration along different contours, this method requires no evaluation of the irregular solutions and therefore is free of any pathology of the charge density near the origin. By explicitly searching the poles of the single-site Green function, we devised an efficient integration scheme to solve the numerical problems caused by the  shallow bound states and the resonance states. The density of states and bulk modulus of polonium are calculated, with the lattice constant found to be $a=3.327$ {\AA}, and the bulk modulus $B=55.1$ GPa, which yield excellent agreement with experimental data and results from other methods. As a test of our code, we also calculated the DOS and bulk modulus of Cu, Ag, and Au, and discussed the impact of relativistic effects.

\section{Acknowledgements}
This work was sponsored by the U.S. Department of Energy, Office
of Science, Basic Energy Sciences, Material Sciences and Engineering
Division. This research used the Extreme Science and Engineering Discovery Environment (XSEDE), which is supported by National Science Foundation grant number OCI-1053575. Specifically, it used the Bridges system, which is supported by NSF award number ACI-1445606, at the Pittsburgh Supercomputing Center (PSC). This research also used the resources of the Oak Ridge Leadership Computing Facility at the Oak Ridge National Laboratory, which is supported by the Office of Science of the U.S. Department of Energy.

\appendix
\section{Pole searching technique} \label{pole}
The pole searching technique is similar to the one used in the quadratic KKR (QKKR) method to quickly finding the poles of the KKR matrix \cite{Sam} corresponding to the energy eigenvalues of the electronic band structure. In scattering theory, the bound states and resonance states correspond to the poles of the S-matrix $\underline{S}(E)$ on complex energy plane, which can be written as
\begin{align}
\underline{S}(E)=\left[-i\underline{s}(E)-\underline{c}(E)\right] \left[i\underline{s}(E)-\underline{c}(E)\right]^{-1},
\end{align}
where $\underline{s}(E)$ and $\underline{c}(E)$ are the sine and cosine matrices\cite{Xianglin}. To find the poles of S-matrix we only need to identify the zeros of the Jost matrix $\underline{J}(E)$, which is given by
\begin{align}
\underline{J}(E)=i\underline{s}(E)-\underline{c}(E).
\end{align}
In scattering theory, the Jost matrix is actually a more fundamental quantity than the S matrix because it has no redundant zeros \cite{arxiv}. To efficiently determine the zeros of the Jost matrix, a linear algebra method is used. Let us consider a square matrix $\underline{J}(z)$ of size $L\times L$ (in relativistic case, $L=2(l_{max}+1)^2$), and we need to find its zeros, $z_p$, such that
\begin{align}
\mathrm{det}\left[ \underline{J}(z_p)\right]=0.
\end{align}
For the present purposes we are only interested in the poles, $\epsilon_p$, that are close to the real energy axis corresponding to either resonance states $(\epsilon_p>0)$ in the valence band or the shallow bound states $(\epsilon_p<0)$. Accordingly, we choose an energy, in the neighborhood of a pole and perform a quadratic expansion of the matrix around $\epsilon_0$ as follows,
\begin{align}
\ud{J}(z)=\ud{J}(\epsilon_0 + \lambda) = \ud{J}^{(0)}(\epsilon_0)+\ud{J}^{(1)}(\epsilon_0) \lambda+
\ud{J}^{(2)}(\epsilon_0) \lambda^2, \label{Jost}
\end{align}
where $\lambda=z-\epsilon_0$. In analogy with the terminology used in the quadratic KKR method, we refer to $\epsilon_0$ as the \textit{pivot energy} for the expansion of $\ud{J}(z)$. To find the zeros of this quadratic equation, we consider an alternative matrix,
\begin{align}
\ud{A}(\lambda)=\left[ J^{(2)}(\epsilon_0)\right]^{-1} \ud{J}(\epsilon_0 + \lambda).
\end{align}
By multiplying $\left[ \ud{J}^{(2)} (\epsilon_0)\right]^{-1}$ on both sides of equation (\ref{Jost}), we get the following expansion,
\begin{align}
\ud{A}(\lambda) = \lambda^2-\ud{B}\lambda + \ud{C},\label{AA}
\end{align}
where
\begin{align}
\ud{B}=-\left[ \ud{J}^{(2)} (\epsilon_0)\right]^{-1} \ud{J}^{(1)}(\epsilon); \;\;
\ud{C}=\left[ \ud{J}^{(2)} (\epsilon_0)\right]^{-1} \ud{J}^{(0)}(\epsilon),
\end{align}
Obviously, $\ud{A}(\lambda)$ has zeros at the same energies,  $\lambda_p = \epsilon_p -\epsilon_0$, as does $\ud{J}(\epsilon_0 + \lambda)$. We now rewrite equation (\ref{AA}) as follows
\begin{align}
\ud{A}(\lambda) = \lambda^2-\ud{B}(\lambda-\ud{D}),
\end{align}
where
\begin{align}
\ud{D}=\ud{B}^{-1} \ud{C} = -\left[ \ud{J}^{(1)} (\epsilon_0)\right]^{-1} \ud{J}^{(0)}(\epsilon).
\end{align}
The zeros of $\ud{A}(\lambda)$ can now be found using the determinantal equation\begin{align}
\mathrm{det}\left[ \ud{A}(\lambda)\right] = \mathrm{det}\left[ \lambda^2-\ud{B}(\lambda-\ud{D})\right]=0,
\end{align}
or, equivalently,
\begin{align}
\mathrm{det}
\begin{bmatrix}
\lambda-\ud{D} & -\ud{I}\\
\ud{D}^2 & \lambda-\ud{B}+\ud{D}
\end{bmatrix}
=0.
\end{align}
The eigenvalues $\lambda_p$ (p = 1, 2,…2L) that satisfy this secular equation can be quickly found by diagonalizing the following matrix:
\begin{align}
\begin{bmatrix}
\ud{D} & \ud{I}\\
-\ud{D}^2 & \ud{B}-\ud{D}
\end{bmatrix}.
\end{align}
The zeros of matrix S(z) are thus at $z_p=\lambda_p+\epsilon_0$.
The zeros of matrix S(z) are thus at $z_p=\lambda_p+\epsilon_0$.
In practice, a simple way of computing the quadratic expansion coefficient matrices in equation (\ref{Jost}) is to calculate $\ud{J}(z)$ at three energy values: $\epsilon_0-\lambda,\epsilon_0,\epsilon+\lambda$, with $\lambda$ an small value, then solve the quadratic expansion equations. To search the poles over the full interval $E_b$ and $E_F$, a set of panels on the real energy axis is set up and $\epsilon_0$ is chosen at the center of each panel to obtain a first approximation to $\epsilon_p$. The accuracy of the pole location can be systematically improved through iteration by progressively decreasing the energy window around the pivot energy used to set up the pole location eigenvalue equation. 

\bibliographystyle{unsrt}
\bibliography{ref01}

\begin{thebibliography}{10}

\bibitem{Korringa}
J.~Korringa.
\newblock {\em Physica}, 13:392, 1947.

\bibitem{Kohn}
W.~Kohn and N.~Rostoker.
\newblock {\em Phys. Rev.}, 94:1111, 1953.

\bibitem{Defects}
P.H. Dederichs, T.~Hoshino, B.~Drittler, K.~Abraham, and R.~Zeller.
\newblock {\em Physica B: Condensed Matter}, 172:203, 1991.

\bibitem{Soven}
P.~Soven.
\newblock {\em Phys. Rev.}, 156:809, 1967.

\bibitem{Gyorffy}
B.~L. Gyorffy.
\newblock {\em Phys. Rev. B}, 1:3290, 1970.

\bibitem{StocksCPA}
G.~M. Stocks, R.~W. Williams, and J.~S. Faulkner.
\newblock {\em Phys. Rev. B}, 4:4390, 1971.

\bibitem{JohnsonPRL}
D.~D. Johnson, D.~M. Nicholson, F.~J. Pinski, B.~L. Gyorffy, and G.~M. Stocks.
\newblock {\em Phys. Rev. Lett.}, 56:2088, 1986.

\bibitem{JohnsonPRB}
D.~D. Johnson, D.~M. Nicholson, F.~J. Pinski, B.~L. Györffy, and G.~M. Stocks.
\newblock {\em Phys. Rev. B}, 41:9701, 1990.

\bibitem{DLM}
J.~Staunton, B.~L. Gyorffy, A.~J. Pindor, G.~M. Stocks, and H.~Winter.
\newblock {\em J. Phys. F: Met. Phys.}, 15:1387, 1985.

\bibitem{GW}
A.~Ernst and M.~L{\"u}ders.
\newblock {\em Methods for band structure calculations in solids}.
\newblock Springer: Berlin, 2004.

\bibitem{DMFT}
J.~Min{\'a}r, L.~Chioncel, A.~Perlov, H.~Ebert, M.~I. Katsnelson, and A.~I.
  Lichtenstein.
\newblock {\em Phys. Rev. B}, 72:045125, 2005.

\bibitem{LSMS}
Y.~Wang, G.~M. Stocks, A.~W. Shelton, D.~M.~C. Nicholson, Z.~Szotek, and W.~M.
  Temmerman.
\newblock {\em Phys. Rev. Lett.}, 75:2867, 1995.

\bibitem{Markus}
M.~Eisenbach, J.~Larkin, J.~Lutjens, S.~Rennich, and J.~H. Rogers.
\newblock {\em Comput. Phys. Comm.}, 211:2, 2017.

\bibitem{KohnSham}
W.~Kohn and L.~J. Sham.
\newblock {\em Phys. Rev.}, 140:A1133, 1965.

\bibitem{Skyrmion}
S.~M{\"u}hlbauer, B.~Binz, F.~Jonietz, C.~Pfleiderer, A.~Rosch, A.~Neubauer,
  R.~Georgii, and P.~B{\"o}ni.
\newblock {\em Science}, 323:915--919, 2009.

\bibitem{Koelling}
D.~D. Koelling and B.~N. Harmon.
\newblock {\em J. Phys. C: Solid State Phys.}, 10.16:3107, 1977.

\bibitem{Tamura}
E.~Tamura.
\newblock {\em Phys. Rev. B}, 45:3271, 1992.

\bibitem{Lovatt}
S.~C. Lovatt, B.~L. Gyorffy, and G.-Y. Guo.
\newblock {\em J. Phys.:Condens. Matter}, 5:8005, 1993.

\bibitem{XDWang}
X.~Wang, X.-G. Zhang, W.~H. Butler, G.~M. Stocks, and B.~N. Harmon.
\newblock {\em Phys. Rev. B}, 46:9352, 1993.

\bibitem{Huhne}
T.~Huhne, C.~Zecha, H.~Ebert, P.~H. Dederichs, and R.~Zeller.
\newblock {\em Phys. Rev. B}, 58:10236, 1998.

\bibitem{Geilhufe}
M.~Geilhufe, S.~Achilles, M.~A. K{\"o}bis, M.~Arnold, I.~Mertig, W.~Hergert,
  and A.~Ernst.
\newblock {\em J. Phys.:Condens. Matter}, 27:435202, 2015.

\bibitem{RLS}
H.~Ebert, J.~Braun, D.~Ködderitzsch, and S.~Mankovsky.
\newblock {\em Phys. Rev. B}, 93:075145, 2016.

\bibitem{Xianglin}
X.~Liu, Y.~Wang, M.~Eisenbach, and G.~M. Stocks.
\newblock {\em J. Phys.: Condens. Matter}, 28:355501, 2016.

\bibitem{Rusanu}
A.~Rusanu, G.M. Stocks, Y.~Wang, and J.~S. Faulkner.
\newblock {\em Phys. Rev. B}, 84:035102, 2011.

\bibitem{Akai}
M.~Ogura and H.~Akai.
\newblock {\em J. Phys.: Condens. Matter}, 17:5741, 2005.

\bibitem{Zeller}
R.~Zeller.
\newblock {\em Phys. Status Solidi b}, 251:2048–54, 2014.

\bibitem{Vladimir}
V.~I. Anisimov, J.~Zaanen, and O.K. Andersen.
\newblock {\em Phys. Rev. B}, 44:943, 1991.

\bibitem{Perdew}
J.~P. Perdew and A.~Zunger.
\newblock {\em Phys. Rev. B}, 23:1981, 1981.

\bibitem{Luders}
M.~L{\"u}ders \textit{et al}.
\newblock {\em Phys. Rev. B}, 71:205109, 2005.

\bibitem{ContourInt}
R.~Zeller, J.~Deutz, and P.~H. Dederichs.
\newblock {\em Solid State Commun.}, 44:993, 1982.

\bibitem{Elastic}
H.~Djohari, F.~Milstein, and D.~Maroudasb.
\newblock {\em Appl. Phys. Lett.}, 90:161910, 2007.

\bibitem{Min}
B.~I. Min, J.~H. Shim, M.~S. Park, K.~Kim, S.~K. Kwon, and S.~J. Youn.
\newblock {\em Phys. Rev. B}, 73:132102, 2006.

\bibitem{Beamer}
W.~H. Beamer and C.~R. Maxwell.
\newblock {\em J. Chem. Phys.}, 14:569, 1946.

\bibitem{Desando}
R.~J. DESANDO and R.~C. LANGE.
\newblock {\em J .inorg. nucl. Chem.}, 28:1837, 1966.

\bibitem{Kraig}
R.~E. Kraig, D.~Roundy, and M.~L. Cohen.
\newblock {\em Solid State Commun.}, 129:411, 2004.

\bibitem{KimPRL}
K.~Kim, H.~C. Choi, and B.~I. Min.
\newblock {\em Phys. Rev. Lett.}, 102:079701, 2009.

\bibitem{LegutPRB}
D.~Legut, M.~Fri{\'a}k, and M.~{\v S}ob.
\newblock {\em Phys. Rev. B}, 81:214118, 2010.

\bibitem{LegutPRL}
D.~Legut, M.~Fri{\'a}k, and M.~{\v S}ob.
\newblock {\em Phys. Rev. Lett.}, 99:016402, 2007.

\bibitem{MinPRB}
C.~Kang, K.~Kim, and B.~I. Min.
\newblock {\em Phys. Rev. B}, 86:054115, 2012.

\bibitem{Lach}
M.~Lach-hab, B.~Akdimand, D.~A. Papaconstantopoulos, M.~J. Mehl, and
  N.~Bernstein.
\newblock {\em J. Phys. Chem. Solids}, 65:1837, 2004.

\bibitem{Rubio}
A.~Rubio-Ponce, J.~Morales, and D.~Olgu{\'i}n.
\newblock {\em Structural and Electronic Properties of Po under Hydrostatic
  Pressure}, pages 119--127.
\newblock Springer Netherlands, 2012.

\bibitem{Papanikolaou}
N.~Papanikolaou, R.~Zeller, and P.~H. Dederichs.
\newblock {\em J. Phys.:Condens. Matter}, 14:2799, 1991.

\bibitem{MacDonald}
A.~H. MacDonald, J.~M. Daams, S.~H. Vosko, and D.~D. Koelling.
\newblock {\em Phys. Rev. B}, 25:713, 1982.

\bibitem{Panaccione}
G.~Panaccione \textit{et al}.
\newblock {\em J. Phys.: Condens. Matter}, 17:2671, 2005.

\bibitem{Strocov}
V.~N.~Strocov \textit{et al}.
\newblock {\em Phys. Rev. Lett.}, 81:4943, 1998.

\bibitem{Marini}
A.~Marini, G.~Onida, and R.~D. Sole.
\newblock {\em Phys. Rev. Lett.}, 88:016403, 2002.

\bibitem{Sam}
J.~S. Faulkner and T.~P. Beaulac.
\newblock {\em Phys. Rev. B}, 26:1597, 1982.

\bibitem{arxiv}
S.~A. Rakityansky, S.~A. Sofianos, and K.~Amos.
\newblock {\em Nuov. Cim. B}, 111:363, 1996.

\end{thebibliography}

\end{document}